\documentclass{article}
\usepackage{epsfig}
\newcommand{\bfr}{\begin{flushright}}
\newcommand{\efr}{\end{flushright}}
 
\begin{document}
\title{Moduli space metric for maximally-charged dilaton black holes
}
\author{Kiyoshi Shiraishi\\
Akita Junior College, Shimokitade-Sakura, Akita-shi, Akita 010, Japan
}
\date{Nuclear Physics {\bf B402} (1993) pp. 399--410 
}
\maketitle
\begin{abstract}
The system consisting of slowly-moving, maximally charged, nonrotating
dilaton black holes is investigated. We obtain the metric on the moduli
space of the system in the low-velocity limit. We find that: (1) only
two-body interactions exist between the extreme black holes in string
theory; (2) the mutual interaction between the black holes vanishes if
the dilaton can be interpreted as a Kaluza-Klein scalar; and (3) for
general dilaton couplings, there exist many-body interactions among the
extreme black holes. We analyze the low-energy classical scattering of
the two extreme black holes in string theory by utilizing the moduli
space metric.
\end{abstract}

\section{Introduction}
Recently, it is very fashionable to study the properties of exact
soliton solutions in gravitating systems where dilaton couplings exist.
A context of the investigation originates from the study of string
theory \cite{1}. In this case, the dilaton coupling and the other
interactions in the effective field theory are readily determined by
the string theory. On the other hand, there have been many works on the
study of charged black holes in the theory with arbitrary dilaton
coupling \cite{2,3,4,5,6,7}. Through the study on the properties of
black holes, we have learned the fact that the value of dilaton
coupling in string theory (and possibly in the case with no dilaton
coupling
\cite{3,4}) is a critical value for thermodynamics of black holes.

The static multi-black-hole solution in the dilaton-coupled
Einstein-Maxwell system has been obtained by the present author in ref.
\cite{7}. This solution is a generalization of the
Papapetrou-Majumdar-Myers solution \cite{8}. We found the balance of
static forces in the multi-black-hole system described by the solution.
The forces are the Coulomb repulsive force, gravitational force, and
attractive scalar force. In ref.~\cite{7}, we have studied the
interaction between the slowly-moving extremal holes using the
Lienard-Wiechert method at large distances \cite{9,10,11}. We have
found that a special value of the dilaton coupling reduces the total
force to nothing.

In the present paper, we study the motion of the maximally charged
dilaton black holes in the low-velocity limit by calculating an
effective action for the system. The method to obtain the action was
originally developed by Ferrell and Eardley \cite{12}. We manage to
extend their method to the dilaton-coupled system. The effective action
reveals the many-body interaction between the slowly moving extremal
black holes at arbitrary distances. The strength as well as the nature
of the interaction may depend on the value of the dilaton coupling. One
may find the peculiar feature appearing in string theory.

The organization of this paper is as follows.

We show the static multi-centered solution in the $(1+N)$-dimensional
dilatoncoupled system in the subsequent section. Though the exact
solution has already been shown in ref.~\cite{7}, we recapitulate it
here because it is crucial for the calculation in the following
sections.

In sect.~3, we derive the effective action for the multi-black-hole
system of the slow motion. We obtain the exact metric on the moduli
space and discuss the low-energy scattering of two extreme black holes
in string theory in sect.~4. 

The last section is devoted to conclusions.

\section{The multi-dilaton-black-hole solution}
We begin with writing the action for classical fields which we study
here. It takes the following form (without source terms):
\begin{equation}
S=\int d^{N+1}x~\frac{\sqrt{-g}}{16\pi}\left(R-\frac{4}{N-1}(\nabla\phi)^2-e^{-[4a
/(N-1)]\phi} F^2\right)\quad (N\ge 3)\,,
\label{2.1}
\end{equation}
where we fix the Newton constant to unity. The coupling constant $a$
(which can be taken as non-negative) is the parameter which determines
the strength of the coupling between the Maxwell field $F$ and the
dilaton field $\phi$.

For $a=0$, the action is reduced to the one for the usual
Einstein-Maxwell system with a free scalar field. In this case,
multi-centered static solutions to the Einstein-Maxwell equations were
found by Papapetrou and Majumdar and by Myers \cite{8}. Their solutions
describe many-body systems of extremely charged black holes, where the
gravitational attraction and Coulomb repulsion is exactly cancelled in
the static configuration.

For $a=1$, the action reduces to the one which is derived from the
low-energy string theory \cite{1,2,3}. For $a=1$, one can use the
``stringy'' metric $\bar{g}_{\mu\nu}=\exp[4\phi/(N-1)]g_{\mu\nu}$ to get the action
in a familiar form, 
\[
S=\int d^{N+1}x \frac{\sqrt{-g}}{16\pi}
e^{-2\phi}\left[\bar{R}+4(\bar{\nabla}\phi)^2-F^2\right]\,.
\]

The cases with general dilaton couplings are examined in ref.~\cite{7}.
The metric for the $n$-black-hole solution is found to be
\begin{equation}
ds^2=-U^{-2}({\bf x})dt^2+U^{2/(N-2)}({\bf x}) d{\bf x}^2\,, 
\label{2.2}
\end{equation}
where
\begin{equation}
U({\bf x})=(F({\bf x}))^{(N-2)/(N-2+a^2)}\,,
\label{2.3}
\end{equation}
and
\begin{equation}
F({\bf x})=1+\sum_{i=1}^n\frac{\mu_i}{(N-2)|{\bf x}-{\bf x}_i|^{N-2}}\,.
\label{2.4}
\end{equation}

The (Maxwell) vector one-form and the dilaton configuration are
written as 
\begin{equation}
A=\pm\sqrt{\frac{N-1}{2(N-2+a^2)}}(F({\bf x}))^{-1} dt\,, 
\label{2.5}
\end{equation}
and
\begin{equation}
e^{-[4a/(N-1)]\phi}= (F({\bf x}))^{2a^2/(N-2+a^2)}\,.
\label{2.6}
\end{equation}
In this solution, the asymptotic value of $\phi$ is fixed to be zero.

If we set $a=0$ in the solution, we find that the solution reduces
to the Papapetrou-Majumdar-Myers solution in $1+N$ dimensions \cite{8}.

The mass and electric charge corresponding to each point source (=
extreme black hole) is \cite{2,7}
\begin{equation}
m_i=\frac{A_{N-1}(N-1)}{8\pi (N-2+a^2)}\mu_i\,,
\label{2.7}
\end{equation}
\begin{equation}
|Q_i|=\sqrt{\frac{N-1}{2(N-2+a^2)}}\mu_i\,,
\label{2.8}
\end{equation}
where $A_{N-1}=2\pi^{N/2}/\Gamma(N/2)$.

We can define the scalar charge $\sigma$ by examining the asymptotic behavior of
the dilaton \cite{2,7}. The scalar charge associated with each source is
\begin{equation}
|\sigma_i|=\sqrt{\frac{1}{2}(N-1)}\frac{a}{N-2+a^2}\mu_i\,.
\label{2.9}
\end{equation}

Now we can examine the balance in the static forces. The system under
consideration is governed by three kinds of forces; the newtonian attraction, the
Coulomb repulsion, and the attractive scalar force mediated by the dilaton. One
can easily confirm the fact that the total force vanishes in the system
described by the static solution \cite{7}.

We anticipate that an addition of a small amount of kinetic energy to this static
system can be treated by perturbation. In such a case, radiation reactions can be
ignored, because the intensity of the radiations are generally proportional to the
higher order in the velocities. The Lienard-Wiechert potential method has been
applied to the system in ref.~\cite{7}. We exhibit the two-body result
in appendix A in the present paper. By using this method, we know the
properties and strength of the interactions between sources with
arbitrary amounts of charges, but we can hardly get the information of
the interaction at small distances.

In the next section, we apply the method of Ferrell and Eardley
\cite{12} to the system of extreme charged dilaton black holes to get
the information on the slow motion of the holes at arbitrary distances.

\section{The effective action for the many-body system of maximally-charged dilaton
black holes in the low-velocity limit}
In this section we follow the procedure of ref.~\cite{12}, based on the
static solution (\ref{2.2}). We must also be careful not to miss the
contribution from the dilaton field. First we calculate the classical
fields in the presence of the slowly-moving black-hole sources
perturbatively. Second, substituting the perturbative solutions into
the action for the classical field, we obtain the effective action for
maximally-charged dilaton black holes.

First, let us consider the off-diagonal part of the metric and the space components
of the gauge field simultaneously. We need only the $O(v)$ perturbative
solution for these fields to obtain the effective Lagrangian up to 
$O(v^2)$ \cite{12} (Here $v$ represents the velocity of the black hole
as a point source.) The perturbed metric and potential are written in
the form
\begin{equation}
ds^2 =-U^{-2}({\bf x})dt^2+2{\bf N}\cdot d{\bf x} dt+U^{2/(N-2)}{(\bf
x})d{\bf x}^2, 
\label{3.1}
\end{equation}
\begin{equation}
A=\sqrt{\frac{N-1}{2(N-2+a^2)}}(F({\bf x}))^{-1} dt+{\bf A}\cdot d{\bf x}, 
\label{3.2}
\end{equation}
where $U$ and $F({\bf x})$ are defined by (\ref{2.3}) and (\ref{2.4}). We have only
to solve linearized equations with perturbed sources up to $O(v)$ for $N_i$ and
$A_i$. We should note that in the system, the action for changed point sources
(particles) coupled to the dilaton field is formally written as
\begin{equation}
\sum_{i=1}^n\int
ds_i\left(m_ie^{[2a/(N-1)]\phi}+Q_iA_{\mu}\frac{\partial
x^{\mu}_i}{\partial s_i}\right)\,,
\label{3.3}
\end{equation}
which is also checked by looking back on the solution (\ref{2.2}). The coupling
between the vector field and the source does not suffer from the effect of dilaton
coupling.

For notational simplicity as well as gauge invariance, we use the one-forms $P$
and $Q$ which consist of the combination of $N$ and $A$:
\begin{equation}
P=A+\sqrt{\frac{N-1}{2(N-2+a^2)}}
(F({\bf x}))^{-1}Q, 
\label{3.4}
\end{equation}
\begin{equation}
Q=U^2N\,. 
\label{3.5}
\end{equation}

Keeping the first-order terms in $P$ and $Q$, we obtain the similar equations to
those in ref.~\cite{12}. Consequently the solutions are found to be
\begin{equation}
F^{2(a^2-1)/(N-2+a^2)} dP=-\frac{N-a^2}{N-2+a^2}\sum_{i=1}^n Q_i df_i\wedge
{\bf v}_i\cdot d{\bf x}_i\,,
\label{3.6}
\end{equation}
\begin{equation}
F^{(a^2-N)/(N-2+a^2)} dQ=-4\sqrt{\frac{N-1}{2(N-2+a^2)}}\sum_{i=1}^n Q_i df_i\wedge
{\bf v}_i\cdot d{\bf x}_i\,,
\label{3.7}
\end{equation}
where $df$ and ${\bf v}\cdot d{\bf x}$ are one-forms and
\begin{equation}
f_i=\frac{1}{(N-2)|{\bf x}-{\bf x}_i|^{N-2}}\,.
\label{3.8}
\end{equation}

We substitute (\ref{3.6}) and (\ref{3.7}) into the field-theory action with the
boundary term for the Einstein gravity. The second-derivatives of the fields are
eliminated by adding the boundary term. In addition, we must also pay attention to
the perturbation on the dilaton kinetic term and source terms.

Synthesizing all the contributions of low-velocity perturbation, we obtain the
effective Lagrangian up to $O(v^2)$ for $n$ maximally-charged dilaton
black holes,
\begin{eqnarray}
L&=&-\sum_{i=1}^n m_i+\sum_{i=1}^nm_i(v_i)^2+\frac{(N-1)(N-a^2)}{16\pi (N-2+a^2)^2}
\nonumber \\
& &\times\int d^N{\bf x}\,(F({\bf x}))^{2(1-a^2)/(N-2+a^2)}
\sum_{i,j}^n \frac{({\bf n}_j\cdot {\bf n}_j)|{\bf v}_i-{\bf v}_j|^2\mu_i\mu_j}{2
|{\bf r}_i|^{N-1}|{\bf r}_j|^{N-1}}\,,
\label{3.9}
\end{eqnarray}
where ${\bf r}_i={\bf x}-{\bf x}_i$ and ${\bf n}={\bf r}_i/|{\bf r}_i|$.
$F({\bf x})$ is defined by (\ref{2.4}).

Here the background dilaton contribution in the free part of the action is
already regularized. The integration must be carried out with care for the regularization
of divergences and seemingly-divergent integrals.

If we send the dilaton coupling to zero and set $N=3$ in the above
expression, we reproduce the result of ref.~\cite{12}. (The last term
in eq.~(10b) in ref.~\cite{12} has no effect in the integration.)
Furthermore, in the large-separation limit (where we approximate $F(x)$
by one), we find exactly the same result as gained by the
Lienard-Wiechert method
\cite{7} (see appendix A). 

It can directly be observed that the
interaction part of the Lagrangian vanishes when $a^2=N$. This value
of the coupling corresponds to Kaluza-Klein reduction of
$(N+2)$-dimensional space-time to $(N+1)$-dimensional space-time with a
circle ($S^1$)\cite{2}.

A more interesting point we soon become aware of is the existence of many-body
interactions. In general, we obtain infinite species of many-body interactions by
expanding the function F(x). Some special cases arise: (1) when $a^2=0$
and $N=3$, the black holes are governed only by two-body, three-body,
and four-body interactions \cite{12}, (2) when $a^2=0$ and $N=4$,
there are only two-body and three-body interactions, (3) when $a^2=1$
(and any
$N$), there are only two-body interactions, and (4) when $a^2=N$, there
are no interactions.

Therefore in low-energy string theory, there exist only two-body interactions
between maximally-charged black holes, regardless of the dimension of space-time.
This result seems most interesting, and suggests the necessity of further study of
the exact solutions and string theory \cite{1}.

In the next section, we will study the moduli space and the scattering
of two slowly-moving black holes laying particular emphasis on the
string case ($a^2=1$).

\section{The moduli space metric and the slow motion of two maximally
charged black holes in string theory}
We wish to study the slow motion of the maximally-charged dilaton black
holes in the low-energy limit. Here the low-energy limit means the
situation where any radiation reaction can be ignored.

The slow motion of classical lumps or solitons in many kinds of field theoretical
models is expressed by geodesic motion on the moduli space, which is the space of
the parameters in the static configuration \cite{13,14,15}. For the
system consisting of multi-centered extreme Reissner-Nordstrom black
holes, the metric of the moduli space was studied in refs. \cite{10,12}.

In the presence of general scalar interactions, the moduli space metric can be
obtained from the expression (3.9). In this section, we concentrate our attention to
the string case, $a^2=1$.

For $a^2=1$, the interaction part in the Lagrangian becomes very
simple and after integration we get
\begin{equation}
L=-\sum_im_i+\sum_i\frac{1}{2}m_iv_i^2+\frac{A_{N-1}}{16\pi(N-2)}\sum_{i,j}
\frac{|{\bf v}_i-{\bf v}_j|^2\mu_i\mu_j}{2r_{ij}^{N-2}}\,,
\label{4.1}
\end{equation}
where $r_{ij}=|{\bf x}_i-{\bf x}_j|$ and
\begin{equation}
m_i=\frac{A_{N-1}}{8\pi}\mu_i\,. 
\label{4.2}
\end{equation}

Now we focus our attention on the two-black-hole system. The Lagrangian
becomes
\begin{equation}
L_{2B}=-M+
\frac{1}{2}MV^2+\frac{1}{2}\mu
v^2\left(1+\frac{8\pi M}{A_{N-1}(N-2)r^{N-2}}\right)\,, 
\label{4.3}
\end{equation}
where $M=m_a+m_b$,
$\mu=m_am_b/M$, ${\bf V}= (m_a{\bf v}_a +m_b{\bf v}_b)/M$, ${\bf v}={\bf v}_a
-{\bf v}_b$ and
$r=|{\bf x}_a -{\bf x}_b|$.

Thus the metric of the $2N$-dimensional moduli space for the two-body
system turns out to be
\begin{equation}
ds^2=M d{\bf R}^2+\mu\left(1+\frac{8\pi M}{A_{N-1}(N-2)r^{N-2}}\right) d{\bf
r}^2\,, 
\label{4.4}
\end{equation}
where ${\bf R}=(m_a{\bf x}_a + m_b{\bf x}_b)/M$ and ${\bf r}={\bf x}_a -{\bf x}_b$.

The center of mass moves freely, as usually expected. Moreover, we consider a
two-dimensional intersection of the moduli space for simplicity. The two parameters
are the distance between two black holes ($r$) and the azimuthal angle
($\theta$) on the scattering plane.

The global structure of this surface of the moduli space depends on the spatial
dimension $N$. The reduced metric can be written in the form
\begin{equation}
ds^2_{MS}=g(r)(dr^2+r^2d\theta^2)=h(R) dR^2+R^2 d\theta^2\,,
\label{4.5}
\end{equation}
where
\begin{equation}
g(r)=1+\frac{8\pi M}{A_{N-1}(N-2)r^{N-2}}\,,
\label{4.6}
\end{equation}
and $R(r)$ is a new radial coordinate. $\mu$ has been absorbed in the
rescaling of $r$. 

We can realize the surface immersed in the
three-dimensional euclidean space spanned by the coordinates $R$,
$\theta$, and $z(R)$. Such a surface for the case with $N=3$
and $a=0$ was exhibited in
ref.~\cite{12}. The ``throat'' of the surface is located at the distance
where $h(R(r))$ diverges.

In the present $a^2=1$ case, for $N>4$, the throat is located at
\begin{equation}
r=\left(\frac{4\pi(N-4)M}{A_{N-1}(N-2)}\right)^{1/(N-2)}\,. 
\label{4.7}
\end{equation}
The surface looks like fig.~1 of ref.~\cite{12} in this case (fig.~1a).

\begin{figure}[ht]
\begin{center}
\includegraphics[width=6cm]{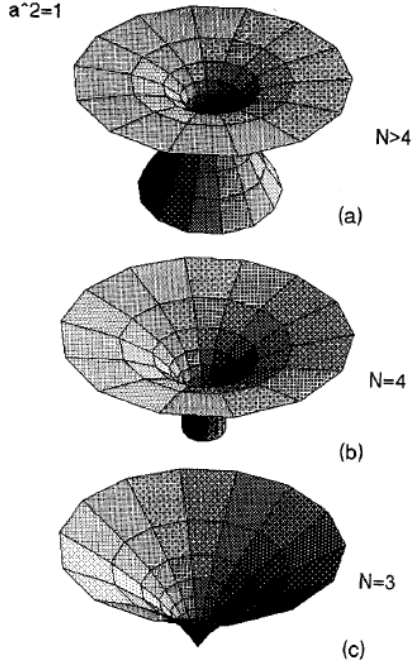}
\caption{A schematic view of the geometry of the moduli space of the two-body
system of extreme black holes in string theory ($a^2 = 1$). (a) for the case $N>
4$, (b) for the case $N=4$, and (c) for the case $N=3$. 
}
\label{f1}\end{center}
\end{figure}

In the $a^2=1$ case, the surface of the moduli does not have a throat
if $N=3$ or $4$. For $N=4$, the surface near the origin looks like a
``tube'', and the origin $r=0$ corresponds to the infinity far down
the tube (fig.~1b).

For $N=3$, the surface has the shape such as fig.~1c. There is a
deficit angle $r$ around the origin of the moduli space for the two-body
problem, while the moduli space in the asymptotic region approaches a
flat space with no deficit angle. Suppose that the two black holes
approach each other from spatial infinity with low relative velocity
and impact parameter $b$. For $N>4$ (and $a^2=1$), the two black holes
coalesce if $b$ is smaller than the critical value
\begin{equation}
b_{coal}=\sqrt{\frac{N-2}{N-4}}\left(\frac{4\pi
(N-4)M}{A_{N-1}(N-2)}\right)^{1/(N-2)}\quad (N>4)\,, 
\label{4.8}
\end{equation}
which is in the same order of the radius of the throat.

For $N=4$ and $a^2=1$, the critical value for $b$ is given by
\begin{equation}
b_{coal}=\sqrt{\frac{2M}{\pi}}\quad (N=4)\,. 
\label{4.9}
\end{equation}

For $N=3$ and $a^2=1$, the black holes never coalesce according to
the geodesic approximation on the moduli space.

\begin{figure}[ht]
\begin{center}
\includegraphics[width=6cm]{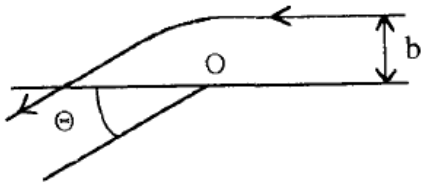}
\caption{The deflection angle $\Theta$.}
\label{f1}\end{center}
\end{figure}

For $N=3$ and $N=4$, the deflection angle (see fig.~2) can be exactly
calculated. The results are the following ($b>b_{coal}$ for $N=4$)
\begin{eqnarray}
\Theta&=&2\tan^{-1}\frac{M}{b}\quad (N=3~\mbox{and}~a^2=1)\,,\\
\label{4.10}
\Theta
&=&\pi\left[\left(1-\frac{2M}{\pi b^2}\right)^{-1/2}-1\right] (b >
b_{coal},~N=4~\mbox{and}~a^2=1)\,.
\label{4.11}
\end{eqnarray}

In the case of $N=3$, the naive limit $b\rightarrow 0$ yields $\Theta=
\pi$, i.e., the backward scattering. This behavior can be derived from
the nature of the metric of the moduli space. (The scattering of two
extreme black holes in four-dimensional string theory appears much
alike that of two BPS monopoles \cite{13,15}; a crucial difference is,
however, the fact that we can tell the distinction of two black holes.)
By now, however, we do not know the applicability of the low-energy
regime to the case with small impact parameter.

The differential cross section in this $N=3$ case turns out to be in
the form of the Rutherford scattering,
\begin{equation}
\frac{d\sigma}{d\Omega}=\frac{1}{4}\frac{M^2}{\sin^4(\frac{1}{2}\Theta)}\,. 
\label{4.12}
\end{equation}

The analysis of higher-dimensional cases and of the case with general dilaton
coupling, which is of much importance, can onlybe done by the help of numerical
calculations. This is beyond the scope of the present paper.

\section{Conclusion}
We have obtained the effective action for the system consisting of $n$
maximally charged dilaton black holes in the low-velocity limit. The
analysis has been performed in $1+N$ dimensions ($N>3$) and is valid
at small distances. 

We have found a peculiar feature appearing in string
theory. Further, the $N=3$ case is a special case in that the two
extreme black holes never merge in the low-velocity regime. We should
also notice that the slow motion of the black holes depends only on the
total mass in the two-body case.

Finally we guess that the exact metric for the moduli space can be obtained by
some other way(s) in the case with particular values of dilaton couplings, such as
$a^2=N$ and $a^2=1$, beyond the low-energy regime. We wish to study
this conjecture in the future.

\section*{Appendix A}
Here we show the low-energy effective Lagrangian \cite{9,10,11} for the
system which consists of two point sources separated at a large
distance with general amounts of masses, electric charges, and scalar
charges. (The earlier version of the preprint of ref.~\cite{7} contains
errors in some coefficients.) The Lagrangian which contains up to $O(v^2)$
terms is written in the form 
\begin{eqnarray}
L&=&\frac{1}{2}m_av_a+\frac{1}{2}m_bv_b+\frac{A_{N-1}}{4\pi(N-2)R_{ab}^{N-2}}
\nonumber  \\
&
&\times\left[Q_aQ_b-\sigma_a\sigma_b-\left(\frac{4\pi}{A_{N-1}}\right)^2\frac{2(N-2)}{N-1}
m_am_b\right]
+\frac{A_{N-1}(v_a^2+v_b^2)}{8\pi(N-2)R_{ab}^{N-2}}\nonumber \\
&
&\times\left[-\sigma_a\sigma_b+\left(\frac{4\pi}{A_{N-1}}\right)^2\frac{2N}{N-1}
m_am_b\right] +\frac{A_{N-1}({\bf v}_a\cdot {\bf v}_b)}{8\pi(N-2)R_{ab}^{N-2}}
\left[Q_aQ_b+\sigma_a\sigma_b\right.\nonumber  \\
&
&~~\left.-\left(\frac{4\pi}{A_{N-1}}\right)^2\frac{2(3N-2)}{N-1}
m_am_b\right]
+\frac{A_{N-1}({\bf n}\cdot {\bf v}_b)({\bf n}\cdot {\bf
v}_b)}{8\pi(N-2)R_{ab}^{N-2}}\nonumber  \\
&
&\times\left[(N-2)Q_aQ_b-(N-2)\sigma_a\sigma_b-\left(\frac{4\pi}{A_{N-1}}\right)^2
\frac{2(N-2)^2}{N-1}
m_am_b\right]+\cdot\,,
\label{A.1}
\end{eqnarray}
where $R_{ab}$ is the distance between two point sources denoted $a$ and
$b$, and ${\bf n}$ is the unit vector in the direction $b-a$.

When the relation of the extreme charged dilaton black hole (\ref{2.7}),
(\ref{2.8}), (\ref{2.9}) are substituted, we find
\begin{equation}
L=\frac{1}{2}m_av_a^2+\frac{1}{2}m_bv_b^2+
\frac{A_{N-1}(N-1)(N-a^2) |{\bf v}_a-{\bf v}_b|^2\mu_a\mu_b}{16\pi(N-2)(N-2+a^2)^2
R_{ab}^{N-2}}+\dots\,,
\label{A.2}
\end{equation}
which precisely recovers the result obtained in the text in the large-distance limit
($F$ is replaced by one).



\begin{thebibliography}{99}
\bibitem{1} A. Dabholkar et al., Nucl. Phys. {\bf B340} (1990) 33;

A. Strominger, Nucl. Phys. {\bf B343} (1990) 167;

E. Copeland, D. Haws and M. Hindmarsh, Phys. Rev. {\bf D42} (1990) 726;

M. J. Duff and J. X. Lu, Phys. Rev. Lett. {\bf 66} (1991) 1402; Nucl. Phys.
{\bf B354} (1991) 129, 141; {\bf B357} (1991) 534; Phys. Lett. {\bf B273} (1991)
409;

M. J. Duff and K. S. Stelle, Phys. Lett. {\bf B253} (1991) 113;

R. Guven, Phys. Lett. {\bf B276} (1992) 49;

R. R. Khuri, Phys. Lett. {\bf B259} (1991) 261;

C. G. Callan, Jr. and R. R. Khuri, Phys. Lett. {\bf B261} (1991) 363;

C. G. Callan Jr., J. A. Harvey and A. Strominger, Nucl. Phys. {\bf B359}
(1991) 611; {\bf B367} (1991) 60;

G. T. Horowitz and A. Strominger, Nucl. Phys. {\bf B360} (1991) 197;

J. A. Harvey and A. Strominger, Phys. Rev. Lett. {\bf 66} (1991) 549;

S. B. Giddings and A. Strominger, Phys. Rev. Lett. {\bf 67} (1991) 2930;

D. Brill and G. T. Horowitz, Phys. Lett. {\bf B262} (1991) 437;

I. Martin and A. Restuccia, Phys. Lett. {\bf B271} (1991) 361;

J. Home, G. Horowitz and A. Steif, Phys. Rev. Lett. {\bf 68} (1992) 568;

J. H. Horne and G.T. Horowitz, Nucl. Phys. {\bf B368} (1992) 444;

J.A. Harvey and J. Liu, Phys. Lett {\bf B268} (1991) 40.

\bibitem{2} G.W. Gibbons and K. Maeda, Nucl. Phys. {\bf B298} (1988) 741.

\bibitem{3} D. Garfinkle, G. Horowitz and A. Strominger, Phys. Rev. {\bf D43}
(1991) 3140; {\bf D45} (1992) 3888 (E).

\bibitem{4} J. Preskill, P. Schwarz, A. Shapere, S. Trivedi and F.
Wilczek, Mod. Phys. Lett. {\bf A6} (1991) 2353.

\bibitem{5} C. F. E. Holzhey and F. Wilczek,
Nucl. Phys. {\bf B380} (1992) 447.

\bibitem{6} K. Shiraishi, Phys. Lett. {\bf A166} (1992) 298.

\bibitem{7} K. Shiraishi, J. Math. Phys. {\bf 34} (1993) 1480
(arXiv:1402.5484).

\bibitem{8} A. Papapetrou, Proc. R. Irish Acad. {\bf A51} (1947) 191;

S. D. Majumdar, Phys. Rev. {\bf 72} (1947) 930;

R. C. Myers, Phys. Rev. {\bf D35} (1987) 455.

\bibitem{9} P. J. Ruback, Commun. Math. Phys. {\bf 107} (1986) 93.

\bibitem{10} G. W. Gibbons and P. J. Ruback, Phys. Rev. Lett. {\bf 57} (1986)
1492.

\bibitem{11} L. D. Landau and E. M. Lifschitz, {\it Classical theory of
fields}, 4th ed. (Pergamon, Oxford, 1975).

\bibitem{12} R. C. Ferrell and D. M. Eardley, Phys. Rev. Lett. {\bf 59} (1987)
1617.
\bibitem{13} N. S. Manton, Phys. Lett. {\bf B110} (1982) 54.
\bibitem{14} R. Ward, Phys. Lett. {\bf B158} (1985) 424.
\bibitem{15} M. Atiyah and N.J. Hitchin, {\it The geometry and dynamics of
magnetic monopoles} (Princeton U. P., Princeton, NJ, 1988).
\end{thebibliography}
\end{document}